\documentclass[aps,prl,reprint,superscriptaddress,showkeys]{revtex4-2}

\usepackage{siunitx}
\usepackage[version=4]{mhchem}
\usepackage{cleveref}
\usepackage{graphicx}
\usepackage[percent]{overpic}

\usepackage{filecontents}

\begin{document}

\title{Real-time Electron Solvation Induced by Bursts of Laser-accelerated Protons in Liquid Water}

\author{A. Praßelsperger}
\affiliation{Fakultät für Physik, Ludwig-Maximilians-Universität München, 85748 Garching, Germany}
\affiliation{Centre for Plasma Physics, School of Mathematics and Physics, Queens University Belfast, Belfast BT7 1NN, United Kingdom}

\author{M. Coughlan}
\affiliation{Centre for Plasma Physics, School of Mathematics and Physics, Queens University Belfast, Belfast BT7 1NN, United Kingdom}

\author{N. Breslin}
\affiliation{Centre for Plasma Physics, School of Mathematics and Physics, Queens University Belfast, Belfast BT7 1NN, United Kingdom}

\author{M. Yeung}
\affiliation{Centre for Plasma Physics, School of Mathematics and Physics, Queens University Belfast, Belfast BT7 1NN, United Kingdom}

\author{C. Arthur}
\affiliation{Centre for Plasma Physics, School of Mathematics and Physics, Queens University Belfast, Belfast BT7 1NN, United Kingdom}

\author{H. Donnelly}
\affiliation{Centre for Plasma Physics, School of Mathematics and Physics, Queens University Belfast, Belfast BT7 1NN, United Kingdom}

\author{S. White}
\affiliation{Centre for Plasma Physics, School of Mathematics and Physics, Queens University Belfast, Belfast BT7 1NN, United Kingdom}

\author{M. Afshari}
\affiliation{Centre for Plasma Physics, School of Mathematics and Physics, Queens University Belfast, Belfast BT7 1NN, United Kingdom}

\author{M. Speicher}
\affiliation{Fakultät für Physik, Ludwig-Maximilians-Universität München, 85748 Garching, Germany}

\author{R. Yang}
\affiliation{Fakultät für Physik, Ludwig-Maximilians-Universität München, 85748 Garching, Germany}

\author{B. Villagomez-Bernabe}
\affiliation{The Dalton Cumbria Facility and the School of Chemistry, The University of Manchester, Oxford Rd, Manchester M13 9PL, United Kingdom}

\author{F. J. Currell}
\affiliation{The Dalton Cumbria Facility and the School of Chemistry, The University of Manchester, Oxford Rd, Manchester M13 9PL, United Kingdom}

\author{J. Schreiber}
\affiliation{Fakultät für Physik, Ludwig-Maximilians-Universität München, 85748 Garching, Germany}

\author{B. Dromey}
\email[]{Corresponding author, email: b.dromey@qub.ac.uk}
\affiliation{Centre for Plasma Physics, School of Mathematics and Physics, Queens University Belfast, Belfast BT7 1NN, United Kingdom}

\date{\today}

\begin{abstract}
\noindent Understanding the mechanisms of proton energy deposition in matter and subsequent damage formation is fundamental to radiation science. Here we exploit the picosecond (\SI{e-12}{s}) resolution of laser-driven accelerators to track ultra-fast solvation dynamics for electrons due to proton radiolysis in liquid water (\ce{H2O}). Comparing these results with modelling that assumes initial conditions similar to those found in photolysis reveals that solvation time due to protons is extended by $>$ \SI{20}{ps}. Supported by magneto-hydrodynamic theory this indicates a highly dynamic phase in the immediate aftermath of the proton interaction that is not accounted for in current models. 
\end{abstract}


\maketitle

Ion interactions in matter, and especially in \ce{H2O}, are of interest for a wide range of fields including radiation chemistry, medical physics and technological applications in the nuclear and space industries \cite{Herb1,Schi1,Newh1,Pral1}. Understanding the impact of mechanisms such as track structure formation is essential for predicting long term radiation effects caused by these energetic particles. While the instantaneous processes underpinning proton interactions in matter and the corresponding ionised electron cascade are well understood \cite{Maro1, Ding1, Emfi1, Salv1}, how the resulting dynamics in the excited medium precipitate a return to equilibrium are less clear. This includes the transition to radiation chemistry and how it combines with an evolving ionised electron distribution to seed permanent damage site formation crucial in applications such as hadron-therapy. A centrally important species in this evolution, and one that is still the centre of much debate, is the solvated electron and the processes underpinning its formation \cite{Svob1}. To date models have largely assumed that solvation in radiolysis and photolysis (photons) can be considered to be identical \cite{Lave2, Bald1}. However to date this assumption has remained largely untested. This is due primarily to the limited temporal resolution provided by conventional radio-frequency accelerators ($\sim$ \SI{100}{ps}). 

Laser-driven accelerators offer a solution to this problem \cite{Borg1}. Recently \citet{Drom1} have implemented a real-time optical streak for the investigation of laser accelerated ion bursts in matter \cite{Senj1}. This allows experimental observation of ultra-fast phenomena which previously could only be studied theoretically. Here we capitalise on this technique to investigate electron solvation dynamics in the immediate aftermath of high-intense proton irradiation of \ce{H2O} with picosecond time resolution. Supported by modelling and a theoretical foundation this provides a detailed picture about solvation yields which potentially influence the subsequent radiation chemistry. This approach offers a route to establishing a fundamental model for how track structures and their evolution can seed the emerging micro-dosimetry \cite{Kell1}.

\begin{figure}[b!]
\includegraphics[width=\linewidth]{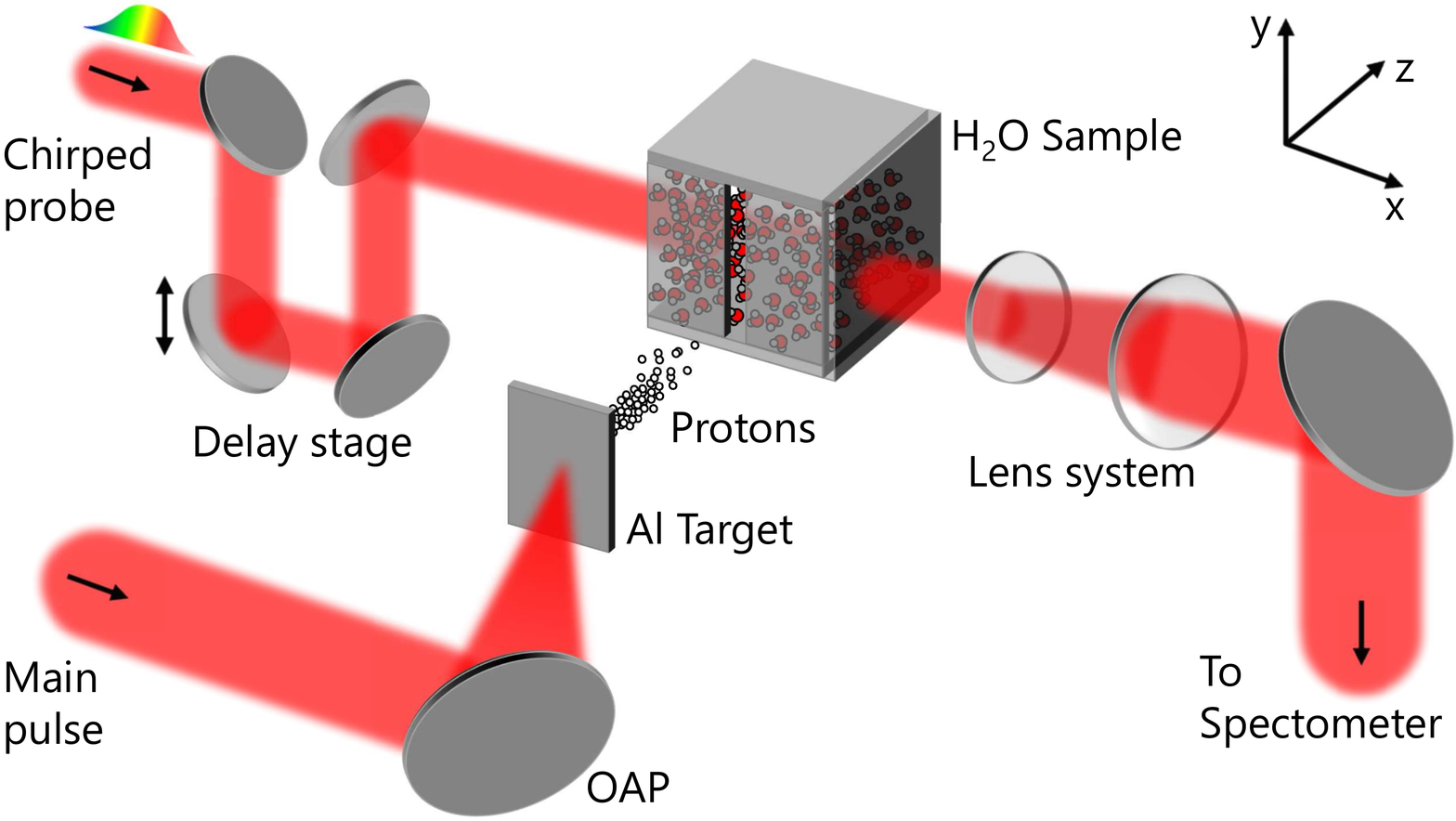}
\caption{\label{fig:setup}\textbf{Scheme of the experimental setting.} The main pulse is focused on the target to produce the proton bunch via TNSA. The protons enter the cell trough a \textit{Teflon}-window. The chirped probe pulse propagates through the sample and is then magnified and introduced into the spectrometer. Depth in the sample is defined in the protons propagation direction (z). Image is not to scale.}
\end{figure}

The experiments were conducted at the \textit{GEMINI} laser facility within the \textit{Rutherford Appleton Laboratory}. The Ti:Sapphire resonators amplify a seed pulse to \SI{15}{J}. The following compression to a temporal FWHM of \SI{30}{fs} and focussing down to the minimum spot size yields a maximum intensity of \SI{2e21}{Wcm^{-2}}. The laser is operating at a central wavelength of \SI{800}{nm}. 

\begin{figure}[t!]
  \begin{overpic}[width=\linewidth]{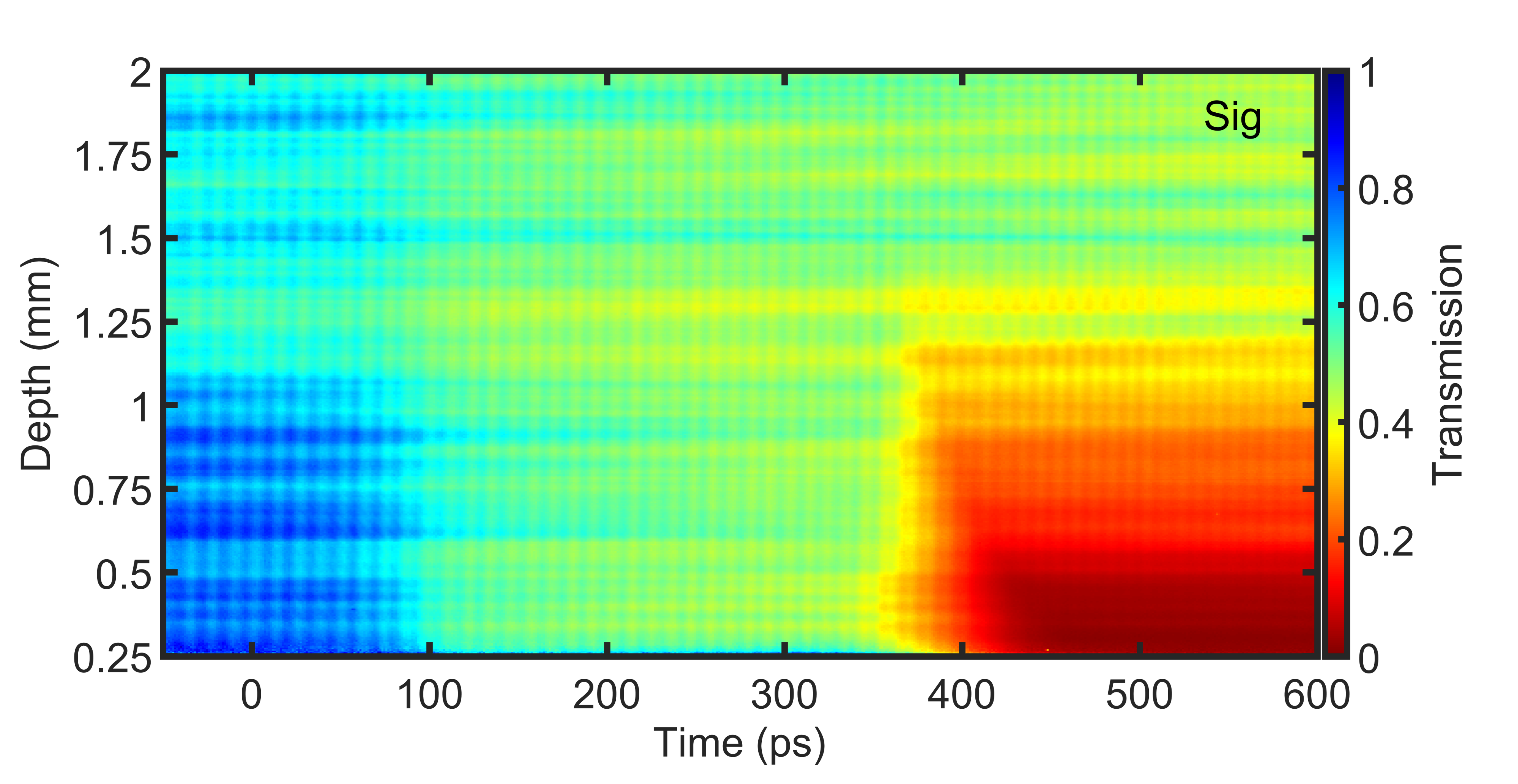}
  \put(0,45){\normalsize\textsf{(a)}}
  \end{overpic}
  \begin{overpic}[width=\linewidth]{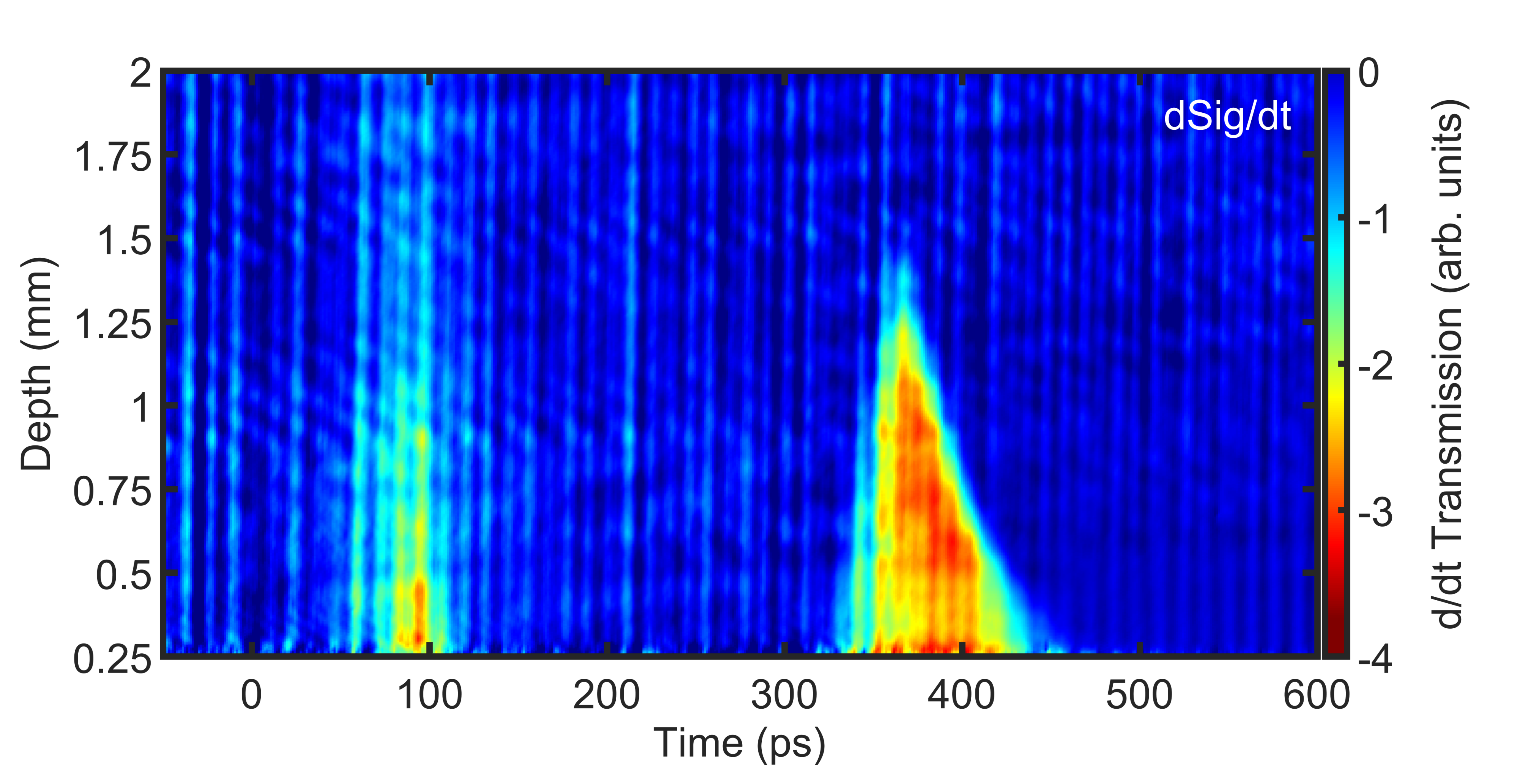}
  \put(0,45){\normalsize\textsf{(b)}}
  \end{overpic}
  \begin{overpic}[width=\linewidth]{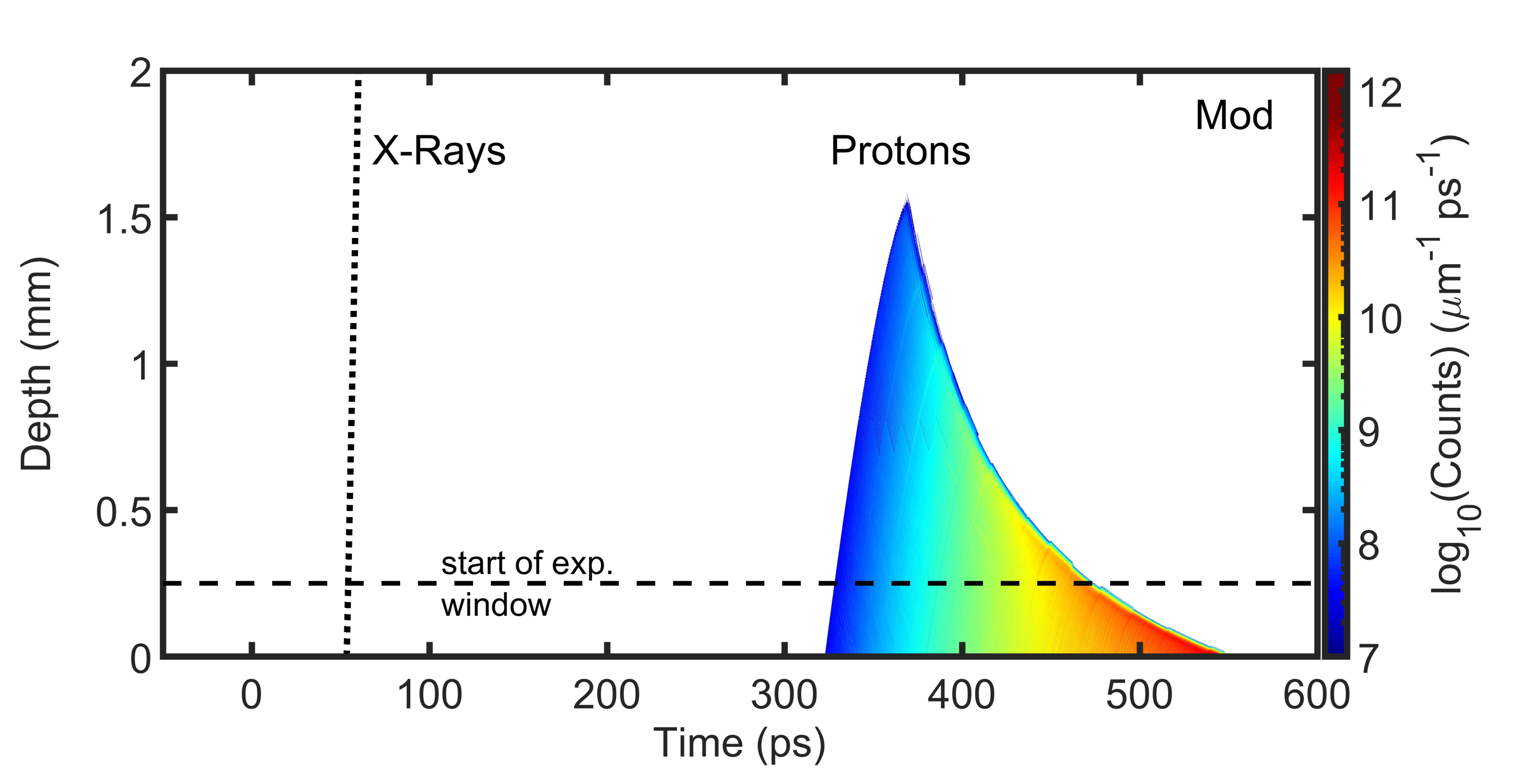}
  \put(0,45){\normalsize\textsf{(c)}}
  \end{overpic}
  \caption{\textbf{Observation of proton interaction with time and depth.} \textbf{(a)} Optical streak of the samples transmission during X-ray and proton interaction in \ce{H2O} (Sig). The maximum incident proton energy was \num{12.3} $\pm$ \SI{0.2}{MeV}. It should be noted that higher energies were available but given that here we are primarily interested in the stopping dynamics for protons interacting in \ce{H2O} energies were chosen to allow this to be studied unambiguously. The interaction of higher energies will be the subject of a future publication.
 \textbf{(b)} Numerical gradient of (a) (dSig/dt). Here the spatio-temporal distribution of the solvation is revealed. Especially the X-ray signal is observed to stop more than \SI{200}{ps} prior to protons' arrival. \textbf{(c)} Computational result of secondary electron counts calculated from reproduction of the experimental proton bunch \cite{Drom1, Coug2} (Mod). In each case, the origin of the time axis corresponds to the point of the X-rays and protons' emission at the target.}
  \label{fig:emap}
\end{figure}

In the experiment the main-pulse was focussed onto a \SI{4}{\mu m} thick aluminium foil under a \SI{40}{^\circ} angle to the target normal using a $f/2$ off-axis parabolic mirror (see \cref{fig:setup}). As fast electrons are accelerated through the target by the driving laser pulse, an initial burst of prompt X-rays generated via bremsstrahlung is emitted. This provides an absolute timing fiducial for the interaction in \ce{H2O} (see \cref{fig:emap}). Next the proton burst is generated by the target normal sheath acceleration (TNSA) mechanism. After their initial acceleration, the proton burst drifts to the water cell containing pristine \ce{H2O}, with a maximum energy of \SI{12.5}{MeV} for these experiments \cite{Roth1, Pass1}. A \SI{500}{\mu m} collimating slit was used to block the off-central part of the proton-beam. The TNSA protons entered the water cell through a \SI{200}{\mu m} \textit{Teflon}-window implying that incident protons with energy less than \num{4.3} $\pm$ \SI{0.1}{MeV} were stopped prior to interacting in the \ce{H2O} sample.

To visualise the proton interaction in the sample a probe-pulse was split from the main pulse. A controllable temporal chirp was introduced to the probe by propagation through a double pass grating set-up. For the experiments discussed here, the temporal FWHM was tuned to approximately \SI{1}{ns}. The synchronisation between both the probe and the main-pulse enabled the exact adjustment of the relative arrival times at the sample and the target, respectively, by a delay stage. The probe beam passed through the proton-\ce{H2O} interaction region transverse to the direction of travel of the TNSA proton bunch. The probe pulse delay was tuned to capture the time-frame of the X-rays and protons interaction in the sample. The on-laser axis region (peak proton energy) of the interaction was magnified and imaged on to the entrance slit of a Czerny-Turner spectrometer with a \SI{10 x 10}{cm}, \SI{1200}{lines/mm} grating. The output from the spectrometer was coupled to a 16-bit CCD camera with \SI{2048 x 2048}{px} on \SI{27.6 x 27.6}{mm} to visualise the results. The linear frequency sweep in time of the chirped probe pulse implies that the temporal evolution of the proton interaction is encoded in its spectrum as a reduction in transmission (see below) as it traverses the interaction region. The chosen chirp delivered a temporal resolution of \SI{1.12}{ps} and the magnification of the probe-beam resolved depth with \SI{4.5}{\mu m} per pixel on the CCD. 

To reduce shot-to-shot fluctuations, two measurements were performed for each proton bunch. That was one probe only reference measurement before the shot and one with the protons interaction. Subsequently the protons signal was divided by the probe only image to provide a normalised spatio-temporal image of the interaction.

The drop in transmission of the probe beam is due to a rapid growth in the radiolytic yield of solvated electrons post irradiation. This is a prototypical species in radiation chemistry in \ce{H2O}. During the solvation process the electrons oscillate between a quasi-free and an excited state with a binding energy of $\approx$ \SI{0.26}{eV} \cite{Pizz1}. This initial oscillation relaxes into the solvated state with a binding energy of $\approx$ \SI{1.5}{eV} \cite{Hert1} on a time-scale of hundreds of femtoseconds \cite{Svob1, Kimu1, Turi1, Yoko1}. In this state the spatial extension of the electron increases, making it possible to gather multiple \ce{H2O} molecules around it and thus increase its binding energy further \cite{Sief1, Abel1, Pizz1}. The central wavelength of the probe beam at \SI{800}{nm} corresponds to a photon energy of \SI{1.55}{eV}. Here absorption is at \SI{80}{\%} of the maximal value \cite{Kimu1, Turi1, Jou1}. The strong coupling can be explained by the superposition of the transition resonances of the quasi-free and the excited state.

Notably, electron thermalisation plus the initial excited state's lifetime are short with respect to the temporal resolution of the experiment. Hence solvation can be expected to happen instantaneously, meaning that around \SI{75}{\%} of all electrons solvate on the pixel that also includes their ionisation. 

A typical optical streak is shown in \cref{fig:emap} (a). The temporal numerical gradient of \cref{fig:emap} (a) is shown in \cref{fig:emap} (b). This is important as it allows the dynamic phase of the interaction to be isolated. In \cref{fig:emap} (a) the long lived nature of the solvated electron means that the drop in transmission of the probe beam persists for \num{10}s of \si{ps} after the initial interaction. By obtaining the gradient of this transmission it reveals the temporal window over which the signal in \cref{fig:emap} (a) is changing and removes the steady state component of the transient absorption after the initial interaction. Due to interference and diffraction effects the range up to \SI{250}{\mu m} was cut in both figures.

Two main features are visible in the data, one starting at \SI{53}{ps} the other one at \SI{326}{ps}. The first signal corresponds to the prompt X-rays \cite{Reth1}. The second feature at \SI{326}{ps} is caused by the proton burst. The proton acceleration process during TNSA is sufficiently short (\SI{30}{fs}) with respect to the temporal resolution of the experiment (\SI{1.12}{ps}), to reasonably assume that X-rays and protons are emitted simultaneously. Thus the maximum proton energy can be estimated by their time-of-flight. For the given data this results in a maximum proton energy of \SI{12.3}{MeV}. An error of \SI{\pm0.2}{MeV} arises from the uncertainty of \SI{\pm50}{\mu m} in detecting the front surface of the sample. This estimate is also corroborated by the stopping range observed for the maximum proton energy in the sample ($\approx$ \SI{1.5}{mm}, \cref{fig:emap}).

It is important at this point to recognise the significance of two key aspects of this experiment that allow for our high accuracy measurements. Firstly, the prompt X-ray pulse provides absolute timing for the experiment, within the uncertainty to which one can measure the source to sample distance. An error of \SI{\pm0.3}{ps} arises, which correlates to one pixel in time. From this all depths and relative times of arrival of the subsequent TNSA proton bunch can be confirmed. Therefore overall uncertainty in the experimental measurement is reduced to noise fluctuations. It is also clear from \cref{fig:emap} (b) that all dynamics due to the interaction of the X-ray pulse have stopped more than \SI{200}{ps} prior to the arrival of the proton bunch. Secondly, the high instantaneous flux of protons of $\approx$ \SI{100}{\mu m^{-2}} in \SI{0.5}{MeV} bandwidth allows the observation of a strong transient absorption signal due to solvated electron generation without the need for scavaging agents \cite{Drom1, Coug2}. This means that using this technique the proton interaction occurs in pristine \ce{H2O}.

In \cref{fig:emap} (c) the simulated proton interaction is shown. The X-ray pathway is indicated by the vertical dotted line, the horizontal dashed line shows the start of the experimental window. The color-map denotes the number of ionised electrons per unit time and depth. The highest considered proton energy was \SI{12.5}{MeV} to reproduce experimental conditions.

It is important to interpret the spatio-temporal profile for the broadband TNSA bunch stopping in \ce{H2O}. The leading edge of the signal with depth corresponds to the propagation path of the highest energy protons in the sample. Energy deposition leads to a deceleration, resulting in a stopping of these protons at approximately \SI{1.5}{mm}. With elapsing time lower energy protons arrive at the front surface of the sample, showing lower penetration depths, returning the characteristic 'shark tooth' profile for the spatio-temporal stopping observed in \cref{fig:emap} (b) and (c).

To quantitatively calculate the cumulative solvated electron concentration $c_{e_{sol}}$ from the ionisation rates, the following equation was applied:
\begin{equation}
c_{e_{sol}} \propto \int_{dt} \int_{dV} \int_{dE}\mu_{e_{sol}} \; v_e \; S_{ion} \;\frac{d\Psi_p}{dE} \;  dE \; dV \; dt
\label{conc}
\end{equation}
Here, the integrated proton flux spectrum $d\Psi_p/dE$ weighted with the ionisation stopping power $S_{ion}$ is proportional to the local ionisation rate per unit volume $V$. The normalised distribution $v_e$ accounts for the part of the temporally and spatially varying electron spectrum that potentially could get solvated. Lastly, $\mu_{e_{sol}}$ describes the solvation yield which is in competition with other decay mechanisms and strongly depends on the local density of ions, electrons and hydronium radicals (\ce{H3O.}). \citet{Svob1} confirmed that the latter acts as a precursor species of the solvated electron. Precisely, the hydrated \ce{H3O.} will spontaneously decay into \ce{H3O+} and \ce{e-} and thus create a gate for the electron to escape its parenting ion and solvate in the bulk medium \cite{Jun1, Onca1, Sobo1}.

Linking this result to the Beer-Lambert law, allows the transmission $\eta$ to be calculated contingent upon the absorbent concentration  via $log_{10}(\eta) = -\epsilon c_{e_{sol}} l$. Experimental values of the molar absorptivity $\epsilon$ are given by \citet{Kimu1} while $l$ relates to the samples dimensions.

Accordingly, transmission along the path of protons of a given initial energy can be derived. Here two effects superimpose. The ionisation stopping power increases with depth to the Bragg-region. On the other hand the flux decreases due to the protons dissipating both temporally and spatially as successively lower energies stop in the medium. These two effects act in opposition on the radiolytic yield of solvated electrons and consequently the overall decrease in transmission with depth is quite low. To quantitatively characterise the processes triggered by the stopping proton bunch according to \cref{conc}, the local electron spectra and \ce{H3O} species densities related to $v_e$ and $\mu_{e_{sol}}$, respectively, have to be determined.

\begin{figure}[t!]
	\includegraphics[width=\linewidth]{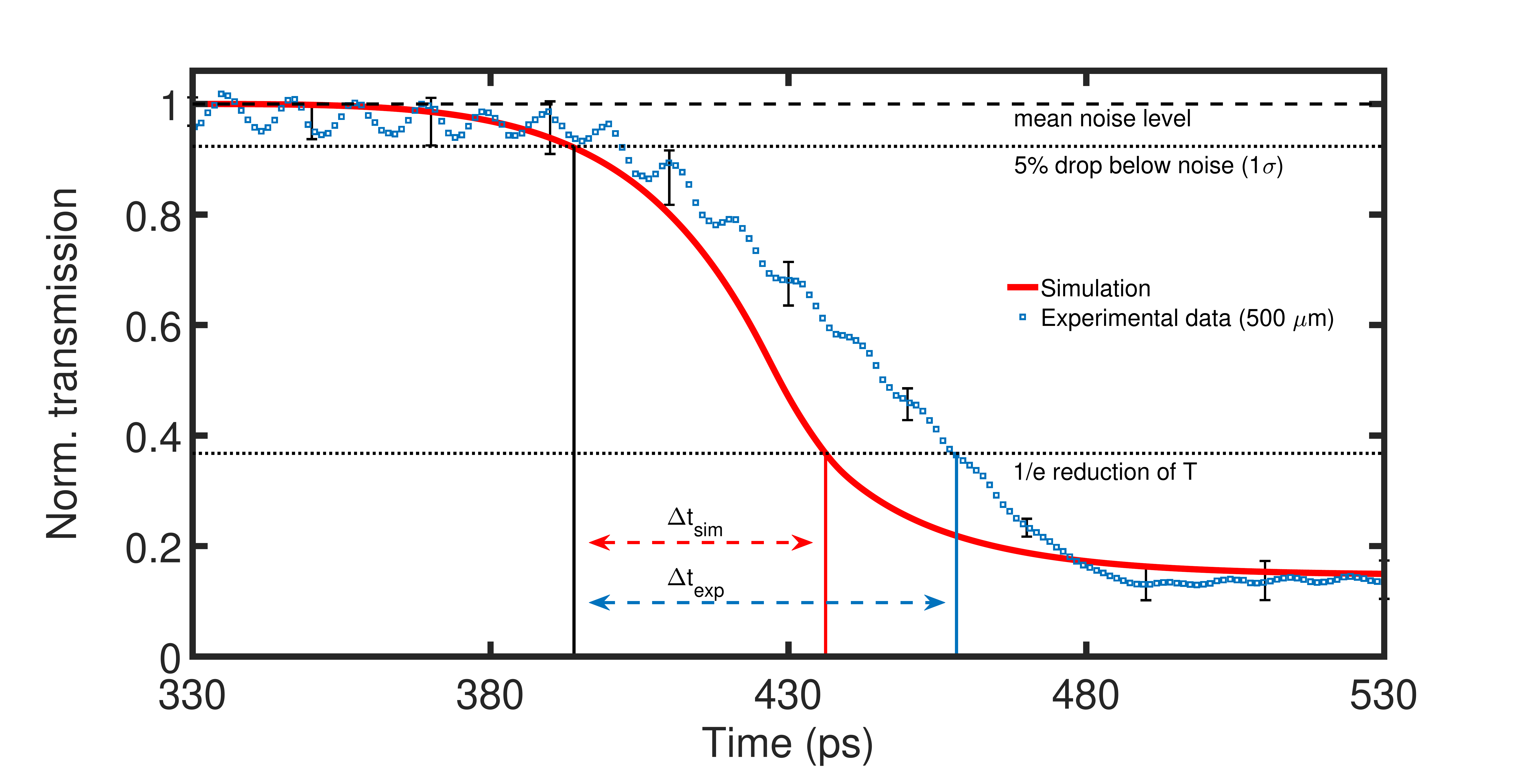}
    \caption{\textbf{Opacity caused by solvated electrons.} Shown is the mean signal of \cref{fig:emap} (a) at \num{500} $\pm$ \SI{50}{\mu m} benchmarked against the transmission caused by the calculated solvation yield. The computation was done for the proton bunch given in \cref{fig:emap} (c). Both line-outs were normalised to \num{1}. The delay of the experimental signal drop with respect to the simulation can be seen by $\Delta t_{exp} - \Delta t_{sim} =$ \num{22} $\pm$ \SI{1}{ps}.}
  \label{fig:esol}
\end{figure}

Therefore, to obtain a complete picture of the temporal evolution of the different molecular, atomic and electronic species in the sample during the protons interaction particle dynamics simulations were performed. This allowed to determine the expected phase-space of a volume within the sample centred at \SI{0.5}{mm} depth. An time dependent energy spectrum of the ionised electrons within the volume was generated to receive a measure for $v_e$ (see \cref{conc}). The emergence and decay dynamics of all particles were tracked by numerical integration of rate equations and therewith $\mu_{e_{sol}}$ (see \cref{conc}) was approximated. The excited solvation state was introduced according to \citet{Svob1}. Based on the density of solvated electrons the probes transmission was computed.

In \cref{fig:esol} the result is plotted against the corresponding line-out of the experiment. Comparing the overall reduction in transmission shows that experimental and modelled concentrations of solvated electrons are in good agreement with each other. However, there is a delayed decline in the experimentally observed dynamics in comparison to that expected from modelling. The time taken for the signal to drop to its $1/e$-value in the experiment is approximately \SI{22}{ps} longer than that calculated in the simulation. This suggests that the solvation process is progressively delayed. 

We find that the temporal discrepancy between the model and the experiment in rooted in the underlying physics based on ultra-fast solvation assumed in the model. From magnetohydrodynamics the macroscopic force acting on a certain density $n_{\alpha}$ of charged particles $\alpha$ can be derived. The movement is described by the centre of mass velocity $\mathbf{u}_{\alpha}$ as:
\begin{equation}
\begin{aligned}
& m_{\alpha} \left( \frac{d (n_{\alpha}\textbf{u}_{\alpha})}{d t} - \frac{d\mathbf{\Gamma}_{\alpha}}{dt} \right) = \\
& q_{\alpha}n_{\alpha}(\textbf{D}+\textbf{u}_{\alpha} \times \textbf{H}) - 3n_{\alpha}\nabla kT_{\alpha} + \textbf{R}_{\alpha \beta}
\end{aligned}
\label{force}
\end{equation}
Here $m$ and $q$ are the particles mass and charge, respectively. The left hand side of this equation describes the plasma's centre of mass movement including its variation by a change in the mass flux $\mathbf{\Gamma}$ due to emerging or decaying particles. The right hand side combines the distinct forces exerted to drive this movement. Here the first term describes the macroscopic field effects $\textbf{D}$ and $\textbf{H}$, respecting also the polarisation and magnetisation of the \ce{H2O} molecules, respectively. The second term contains spatial temperature variations, i.e. the thermal energy $kT_{\alpha}$ drift. The last term $\textbf{R}_{\alpha \beta}$ characterises collisions between the particles and other species $\beta$ which results in macroscopic friction.

As real-time observation of solvated electron formation post proton irradiation was not available prior to the methodology presented here, the majority of data about solvation came from photolysis experiments. Only the friction term $\textbf{R}_{\alpha \beta}$ in \cref{force} can be assumed to be equivalent post proton and photon irradiation as it is dominated by collisions with \ce{H2O} molecules. Differences arise with reference to the other terms. 

Firstly, the incident protons create a nanometre-scale charge reservoir in the Bragg-region as they stop. Here they create a non-equilibrium condition by violating the initial charge-neutrality of the sample. Macroscopic fields build up which corresponds to the $\textbf{D}+\textbf{u}_{\alpha} \times \textbf{H}$ term. The latter will influence the drift especially of \ce{H3O+} ions and electrons contrarily and thus decelerate the solvation yield by separating both species. Additionally it will be highest close to the Bragg-region where the charge surplus assembles which agrees with the observation in \cref{fig:esol}.

Secondly, the transferred energy by the long-ranging Coulomb force raises the average particle energy within the proton tracks drastically. This is best described by the \textit{thermal spike model} \cite{Toul1}. The temperature increase $\Delta T$ in time $t$ and with distance $r$ from the track can be approximated by \cite{Wesc1}:
\begin{equation}
\Delta T(r,t) = \frac{\gamma S}{\pi \rho c a^2(t)}e^{-(r^2/a^2(t))}
\label{temp}
\end{equation}
Here, $\gamma S$ is the deposited energy in thermal spikes and $\rho$ and $c$ are density and heat capacity of \ce{H2O}, respectively. The time dependent factor $a(t)$ describes the dissolution of the spikes. In \ce{H2O} maximum temperature increases of $\Delta T_{max} \approx$ \SI{200}{K} were measured in the Bragg-regime of protons \cite{Toul1}. The temperature gradient generates a lower density around the tracks which reduces the solvation rate (see \cref{force}). Further during the thermal relaxation of the excited solvated state into the ground state, its absorption spectrum shifts from the infra-red to the equilibrium peaked at \SI{721}{nm} \cite{Hert1}. The upper levels absorptivity of \SI{800}{nm} photons is \num{2.8} times lower \cite{Kimu1}. This blue-shift takes hundreds of femtoseconds at \SI{300}{K} \cite{Svob1, Pizz1, Turi1, Yoko1}, however increases to picoseconds at the increased temperatures \cite{Hert1}. Thus absorption gradually delays with increasing temperature.

Both of these effects predominantly drive the charge mobility and, accordingly, the solvation yield $\mu_{e_{sol}}$. This could explain the observed temporal delay of more than \SI{22}{ps} between the simulation and the experiment. This means that the underlying processes of electron solvation in the aftermath of proton bunch presence and passage significantly differ from the ones following photon irradiation.

In summary, we provide direct experimental evidence of ultra-fast electron solvation occurring in the immediate aftermath of proton irradiation of \ce{H2O}. The combination of this real-time observation with modelling revealed a solvation process decelerated by as much as \num{22} $\pm$ \SI{1}{ps}. These results indicate that the dynamics following proton irradiation deviate significantly from the known picture about the solvated state. The underlying physics hints at plasma movements caused by macroscopic fields and temperature spikes to be responsible for this discrepancy.

Relating these fundamental physical processes to long-term radiolytic damages is essential for making high precision predictions. Here the theoretical description and simulation facilitated by real-time optical streaking supply a way to build this model. Especially solutions for the fundamental force-terms (see \cref{force}) and numerical optimisation methods are important to cover a wider range of conditions.

Considering prospective improvements of proton energies \cite{Borg1}, mono-energetic proton bunches \cite{Schw1,Hege1}, means of transporting and focussing these \cite{Scho1,Tonc1} and the growing interest in ion beam applications e.g. by space research \cite{Cuci1}, in ion beam therapy of tumours \cite{Jerm1, Linz1} or even by industry \cite{Hamm1}, this is of uttermost importance. We expect this to contribute a much deeper understanding to physical and chemical micro-dosimetry which precedes an effective use of this rapidly evolving modality.\\
\\

\begin{acknowledgments}
\noindent\textbf{Funding:} The authors acknowledge support by DFG via project GRK2274 and the cluster of excellence Munich Centre for Advanced Photonics. The authors also would like to acknowledge support from EPSRC grants EP/P010059/1 and EP/P016960/1.\\
\textbf{Author contributions:} The experiments were planned and conducted by M.C., N.B., M.Y., C.A., H.D., S.W., M.S., R.Y., F.J.C., J.S. and B.D.. Data were analysed by A.P., M.C., M.A. and B. V.-B.. A.P. performed all the calculations with advice from M.C., M.A., B. V.-B., M.Y., J.S. and B.D.. The draft of the paper was written by A.P. with special support by J.S and B.D.. Improvements followed through contributions from all authors. Collaboration between the Ludwig-Maximilians-Universität München and the Queen's University of Belfast was made possible by ERASMUS+. A.P. also wants to gratefully thank J.S and B.D. for the support during the research project.\\

\end{acknowledgments}

\bibliography{\jobname}

\end{document}